\documentclass[review]{elsarticle}
\def\avg#1{\langle #1\rangle }
\def\bG{{\bf G}}
\def\bk{{\bf k}}
\def\bkh{\bf\hat{k}}
\def\bu{{\bf u}}

\usepackage{amsmath,amssymb}
\usepackage{graphicx}

\usepackage{lineno,hyperref}
\modulolinenumbers[5]

\journal{Materials Chemistry and Physics}

\usepackage{color}

\begin{document}

\begin{frontmatter}

\title{Elastic stability and lattice distortion of refractory high entropy alloys}

\author{Bojun Feng and Michael Widom}
\address{Department of Physics, Carnegie Mellon University}
\address{Pittsburgh, PA  15213}

\begin{abstract}
Refractory high entropy alloys containing elements from the Ti, V and Cr columns of the periodic table form body centered cubic (BCC) structures.  Elements from the Ti column are noteworthy because they take the BCC structure at high temperature but undergo a shear instability and transform to the hexagonal (HCP) structure at low temperature.  We show that the instability of these elements impacts the properties of the HEAs that contain them.  In particular, the shear moduli are greatly reduced, causing increased dynamic contributions to lattice distortion.  Relatively large size differences between elements of the BCC/HCP column compared with the regular BCC columns create additional static contributions to lattice distortion.  These findings are supported by direct evaluation of elastic constants and lattice distortion in four representative HEAs. Comparing moduli of HEAs with those of compositionally averaged pure elements verifies the impact of BCC/HCP elements and allows us to estimate the compositions at which the BCC phases become elastically unstable, predictions that could be useful in material design.
\end{abstract}


\end{frontmatter}


\section{Refractory high entropy alloys}

Many high entropy alloys (HEAs~\cite{Cantor04,Yeh04_1}) contain the refractory elements found in columns 2-4 of the transition metal series in the periodic table~\cite{Senkov10,Senkov11,Maiti16}.  Elements in columns 3 and 4 (those starting with V and Cr, respectively) take a body centered cubic (BCC) structure at all temperatures below melting.  In contrast, those in column 2 (starting with Ti) are BCC at high temperature but transform to hexagonal (HCP) at low temperature through a diffusionless ``martensitic'' transformation known as the Burgers distortion~\cite{Burgers34}.  We refer to these as BCC/HCP elements.  The refractory high entropy alloys, even those containing BCC/HCP elements, form BCC structures as-cast, and have not so far exhibited BCC to HCP transitions at lower temperatures.

The Burgers distortion begins with an orthorhombic shear deformation of the BCC structure~\cite{MasudaJindo04}.  The distortion breaks the symmetry, splits the metal $d$-orbitals and reduces the Fermi level density of states, hence lowering the total energy.  To assess stability against this distortion we consider the elastic moduli of these elements and their alloys.  Cubic structures have three independent elastic constants, $C_{11}$, $C_{12}$ and $C_{44}$ in Voigt notation; there are correspondingly three conditions of elastic stability, the Born rules~\cite{Born40}.  The BCC/HCP elements violate these stability conditions at low temperatures, and they are stabilized at high temperatures through anharmonic vibrational entropy.

\begin{figure}
\includegraphics[trim = 8mm 25mm 40mm 30mm, clip, width=2.5in]{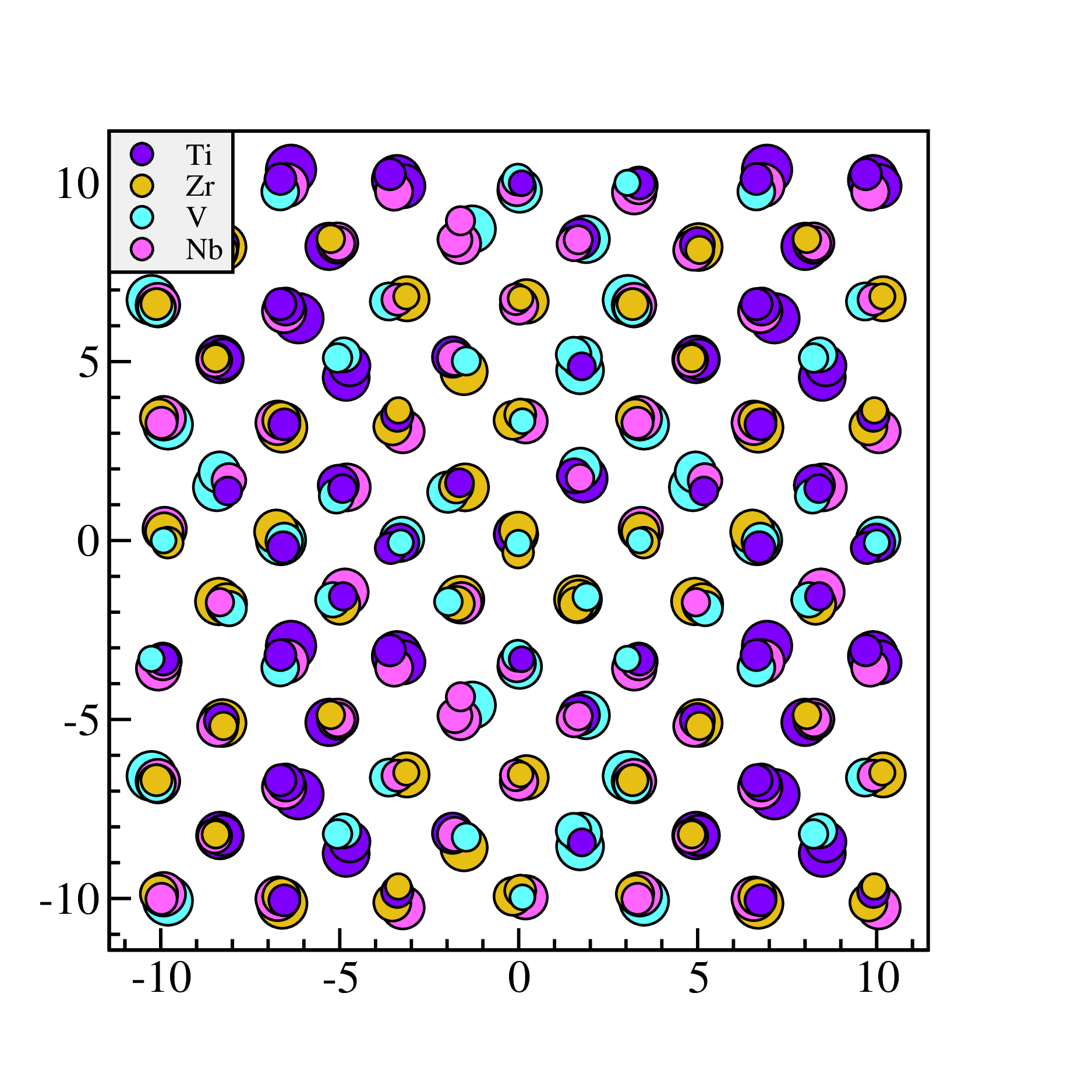}
\includegraphics[trim = 8mm 25mm 40mm 30mm, clip, width=2.5in]{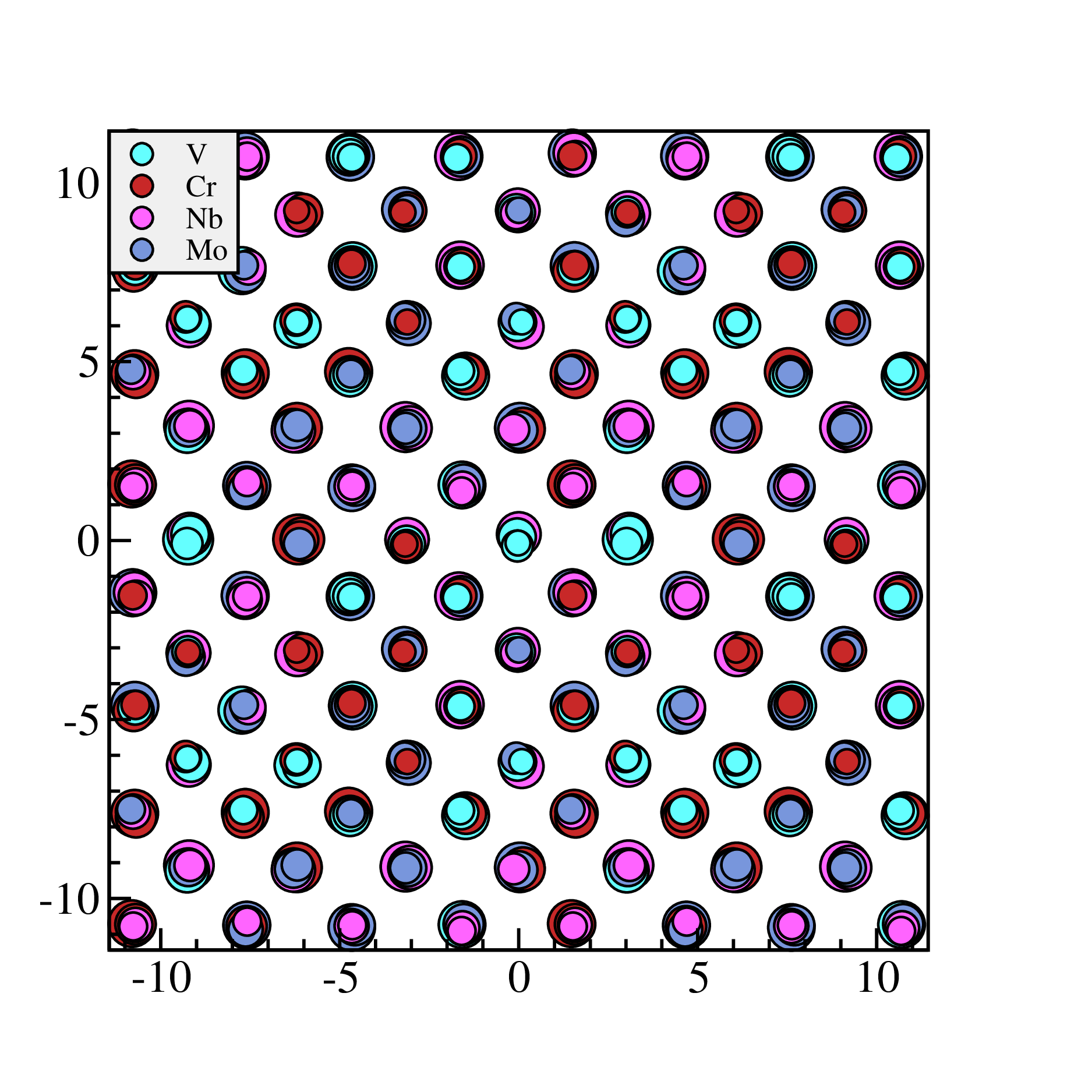}
\includegraphics[trim = 8mm 25mm 40mm 30mm, clip, width=2.5in]{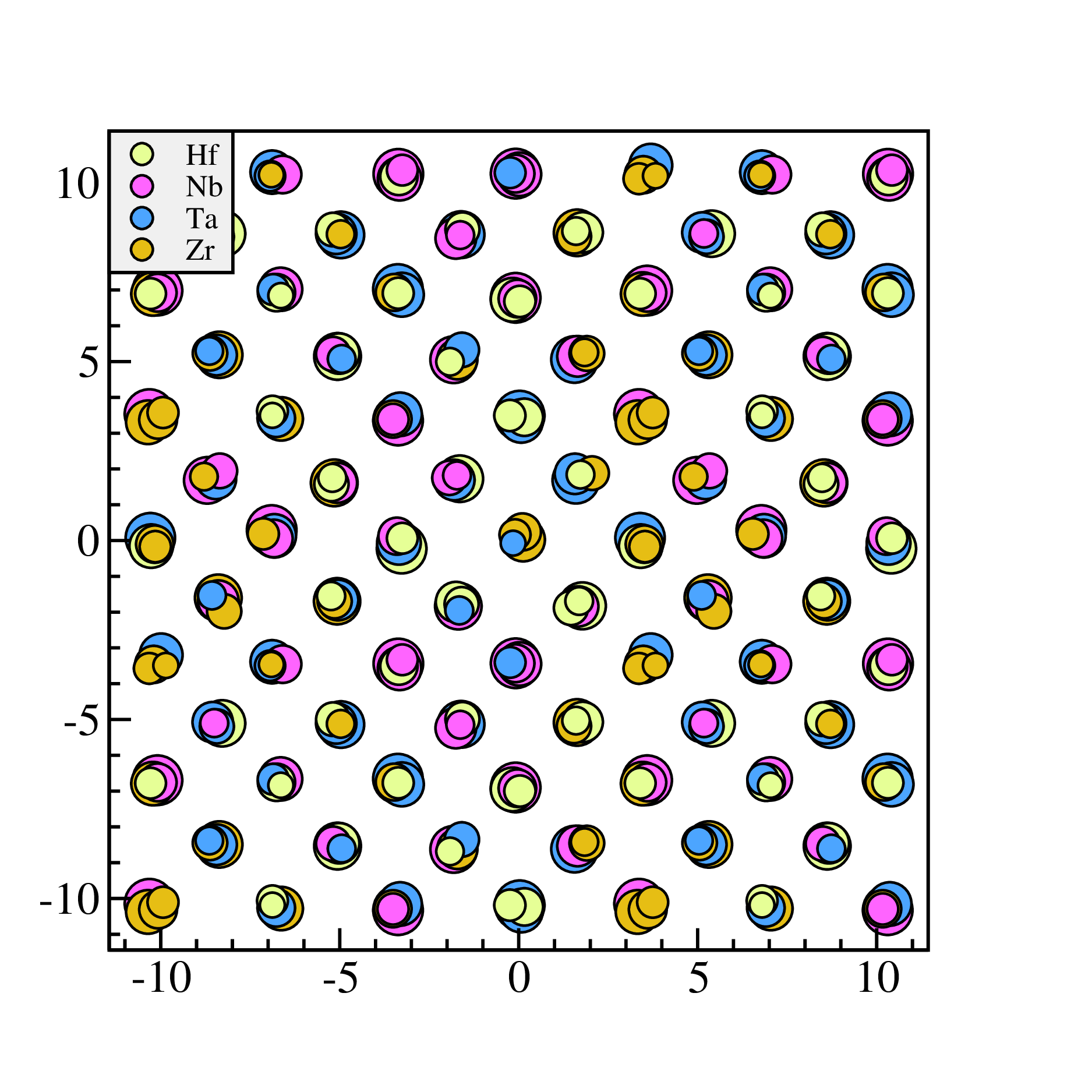}
\includegraphics[trim = 8mm 25mm 40mm 30mm, clip, width=2.5in]{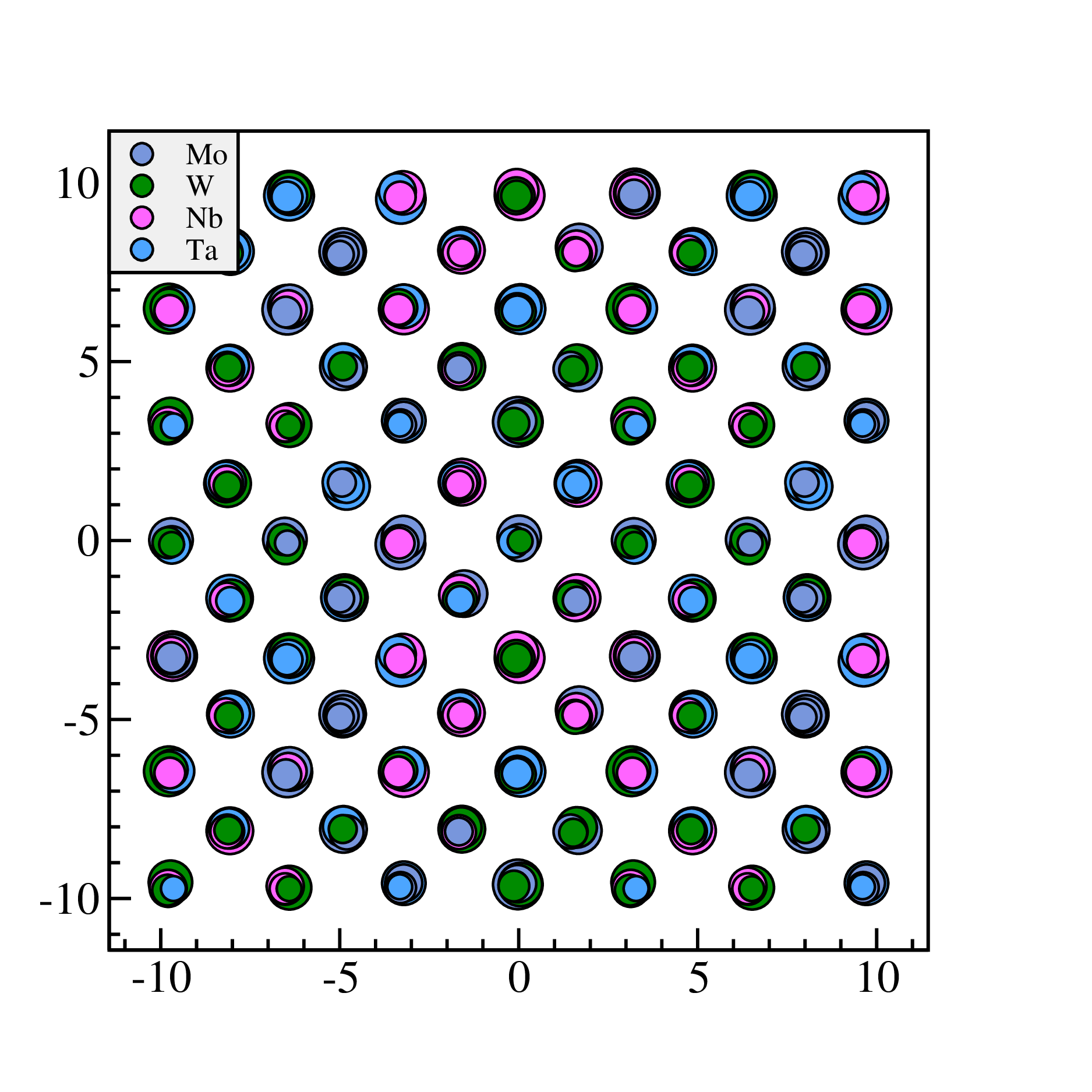}
\caption{\label{fig:structs} Refractory HEA configurations at $T=300$K (quenched from $1200$K).  Atomic species are color coded.  Size indicates vertical height. NbTiVZr and HfNbTaZr contain both BCC/HCP and regular BCC elements, while CrMoNbV and MoNbTaW contain only regular BCC elements.}
\end{figure}

We will explore four representative refractory HEAs comprised of elements drawn from overlapping squares~\cite{Widom16} of the periodic table.  Two of the squares cover the Ti and V columns, and hence contain both BCC/HCP and regular BCC elements.  The other two squares cover the V and Cr columns and hence contain only regular BCC elements.  Fig.~\ref{fig:structs} illustrates representative structures within 128-atom cells simulated at $T=1200$K using hybrid Monte Carlo/molecular dynamics~\cite{Widom13}, then quenched to room temperature $T=300$K using conventional MD, so as to preserve the chemical order characteristic of high temperature. Displacements of atoms off their ideal sites are clearly visible, especially in the alloy systems containing BCC/HCP elements.

\begin{figure}
\includegraphics[trim = 20mm 15mm 20mm 10mm, clip, width=2.0in, angle=-90]{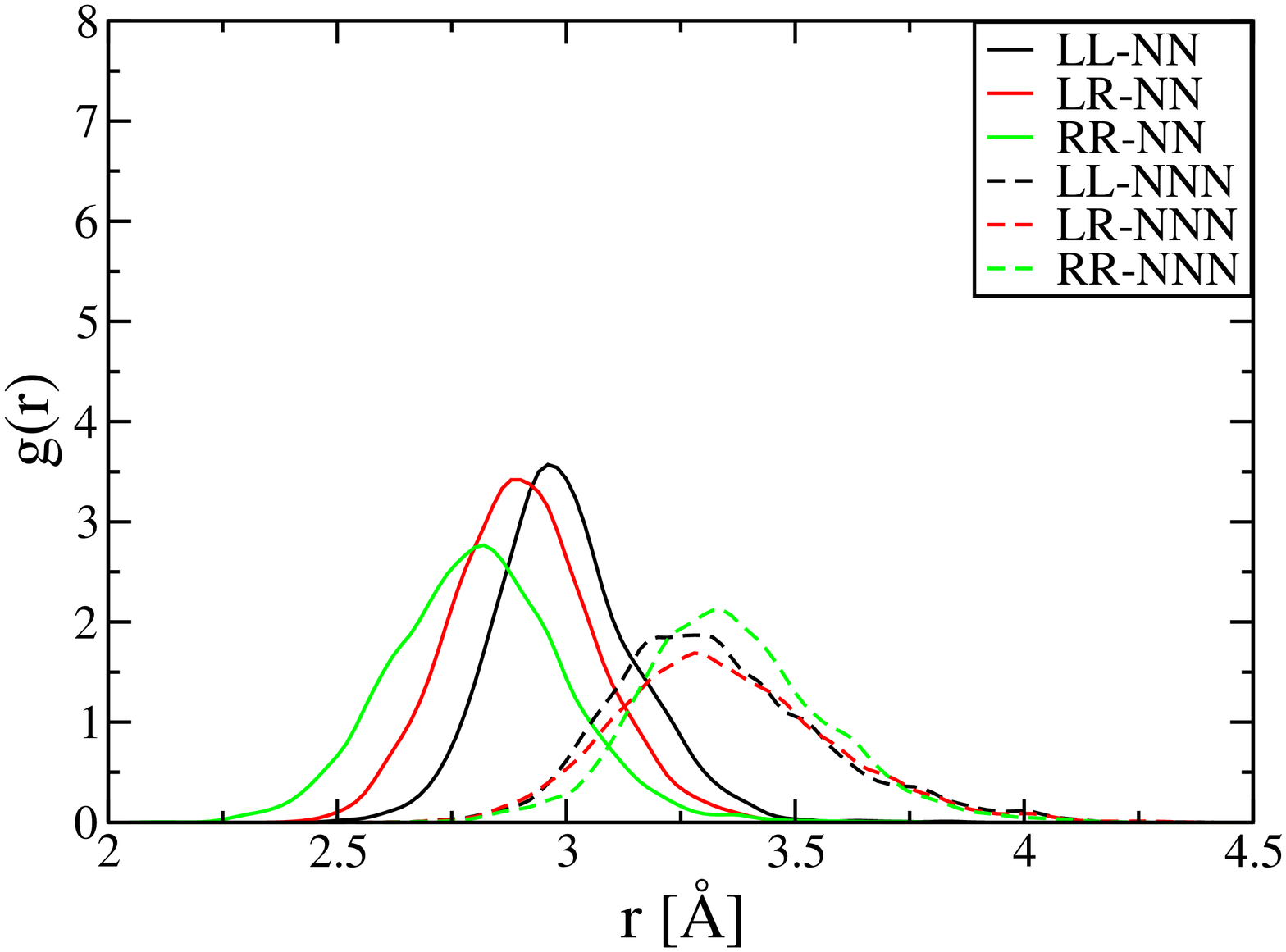}
\includegraphics[trim = 20mm 15mm 20mm 10mm, clip, width=2.0in, angle=-90]{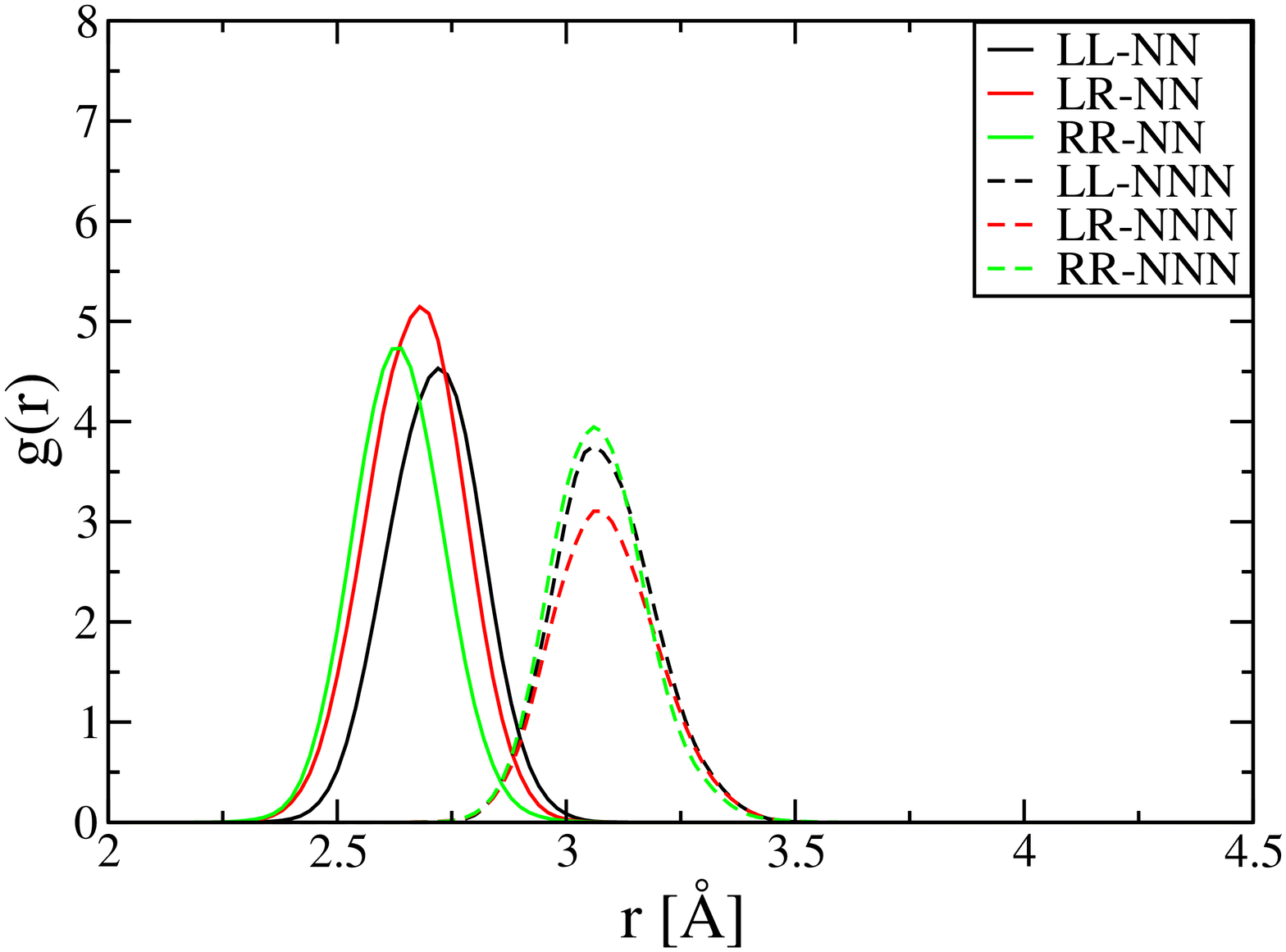}
\includegraphics[trim = 20mm 15mm 20mm 10mm, clip, width=2.0in, angle=-90]{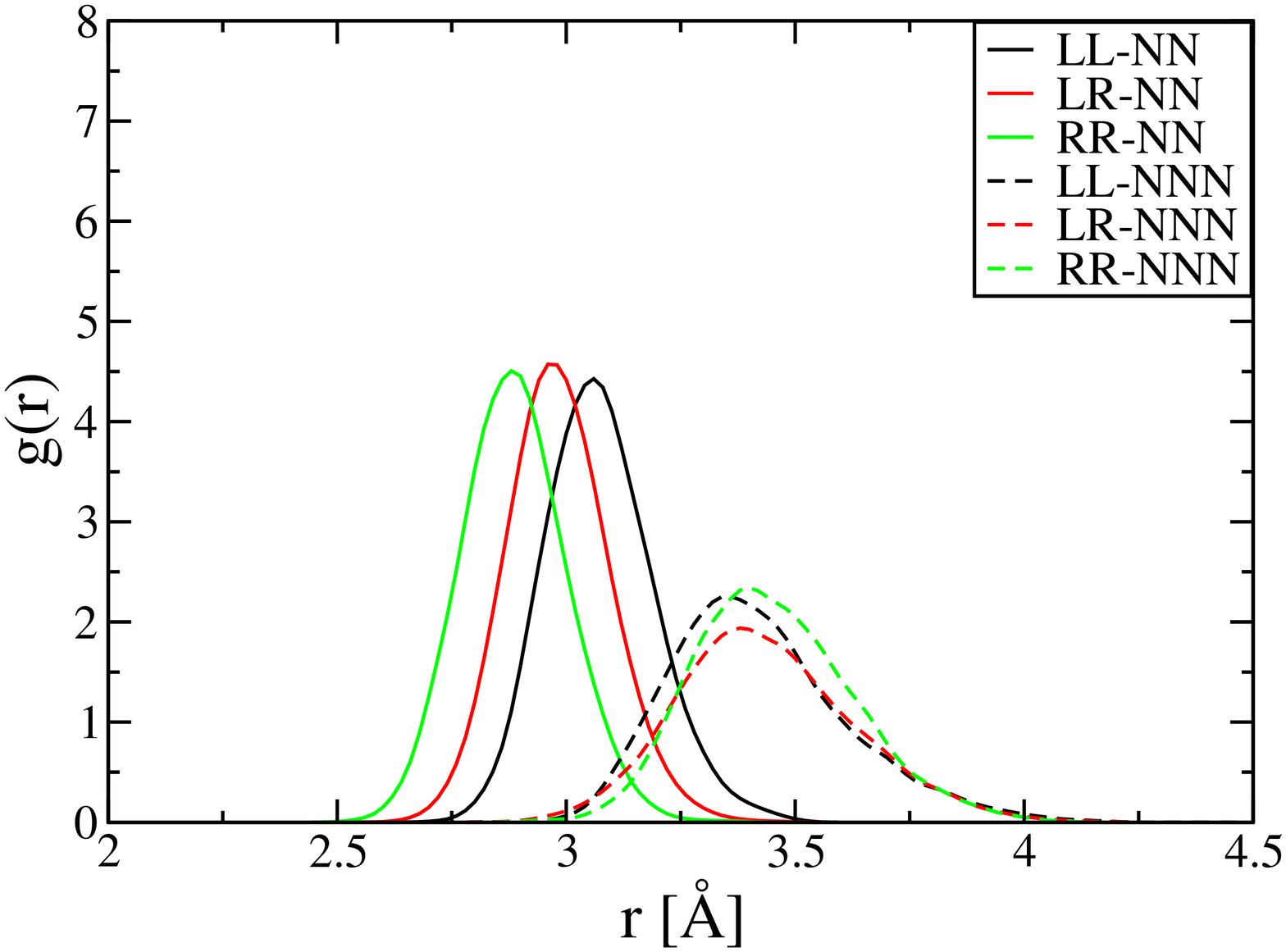}
\includegraphics[trim = 20mm 15mm 20mm 10mm, clip, width=2.0in, angle=-90]{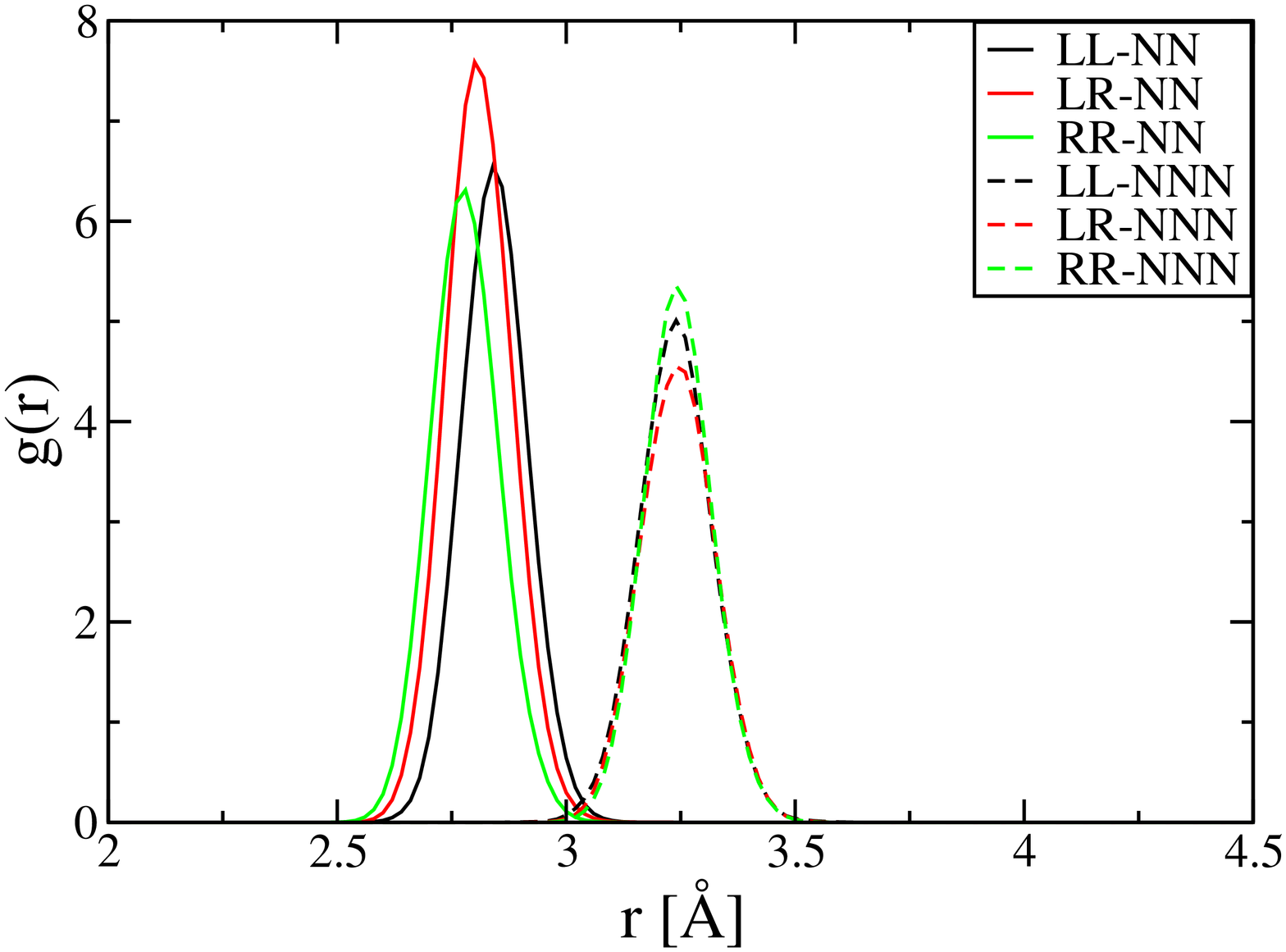}
\caption{\label{fig:pdfs} Partial pair correlation functions averaged over molecular dynamics simulation at $T=300$K. L-L takes elements only from the left-hand periodic table column, L-R are correlations of left-hand and right-hand elements and R-R takes only right-hand elements. For example in NbTiVZr, L-L includes Ti-Ti, Ti-Zr, Zr-Ti and Zr-Zr pairs, while L-R includes Ti-Nb, Ti-V, Zr-Nb and Zr-V, etc. Solid lines are nearest neighbors (NN); dashed are next nearest neighbors (NNN).}
\end{figure}

Pair correlation functions provide an alternate representation of the HEA structure.  Fig.~\ref{fig:pdfs} presents partial pair correlation functions, separated into nearest neighbors (NN) and next nearest neighbors (NNN).  The separation is achieved by defining pairs along the [111] direction as nearest neighbors and along [100] as next nearest.  Rather than display all ten partials~\cite{Widom16}, $g_{\alpha\beta}$, for each alloy system, we group them into three classes according to which columns of the periodic table are represented; L and R indicate the left-hand and right-hand columns respectively.  Several trends are evident.  In all cases the NN correlations occur at larger distances for L-L than L-R with R-R being shortest, in keeping with the periodic table trend of decreasing atomic radius from left to right within each transition metal row.  NN and NNN peaks overlap considerably for the two compounds containing BCC/HCP elements, and the overlaps are greatest for the L-L correlations which are specifically those of the BCC/HCP elements.  In fact, summing the NN and NNN correlations, no separation can be observed between the NN and NNN peaks in these partials, as has been separately noted in the case of NbTiVZr~\cite{Widom16}.  Such effects have been reported experimentally in the medium entropy alloy HfNbZr~\cite{Guo2013}.

\section{Elasticity}

We calculate elastic constants from stress-strain relationships of the two-atom BCC unit cells using numerical two point central differences as implemented in VASP~\cite{Kresse99} in the generalized gradient approximation~\cite{Perdew96}.  Energy cutoffs are increased to 400 eV and $k$-point densities are increased to achieve convergence of 1 GPa on all elastic constants.  Elemental Cr is treated as an antiferromagnet.  Because atomic coordinates and lattice constants are fully relaxed, elastic moduli are predicted for $T=0$K.  

For all structures, we report the elastic moduli $C_{ij}$ and also the Hill average~\cite{Hill52} of the Voigt~\cite{Voigt28} and Reuss~\cite{Reuss29} polycrystalline moduli.  For cubic materials these are~\cite{Jong2015}
\begin{equation}
\label{eq:bulk}
K_V=K_R=K_H=\frac{1}{3}\left(C_{11}+2C_{12}\right)
\end{equation}
and
\begin{equation}
\label{eq:shear}
G_V=\frac{1}{5}\left(C_{11}-C_{12}+C_{44}\right),~~
G_R=\frac{5\left(C_{11}-C_{12}\right)C_{44}}{3C_{11}-3C_{12}+4C_{44}},~~
G_H=(G_V+G_R)/2.
\end{equation}
We also report the Poisson ratio
\begin{equation}
\label{eq:Poisson}
\sigma=(3K_H-2G_H)/(6K_H+2G_H)
\end{equation}
and also the Zener anisotropy~\cite{Zener47} (ratio of shear moduli)
\begin{equation}
\label{eq:aniso}
A_A=C_{44}/\mu=2C_{44}/(C_{11}-C_{12})
\end{equation}
for which $A=1$ indicates isotropy.  In Eq.~(\ref{eq:aniso}) we have defined the shear moduli as $C_{44}$ and
\begin{equation}
\label{eq:mu}
\mu\equiv(C_{11}-C_{12})/2.
\end{equation}
Born rules for stability~\cite{Born40} require positivity of the bulk modulus $K_H$ and the two shear moduli, $C_{44}$ and $\mu$.  These, in turn, imply bounds on the Poisson ratio $-1 < \sigma < 1/2$, and positivity of the anisotropy $A_Z>0$.

\begin{table}
\caption{\label{tab:elements} Calculated elastic constants of elemental BCC refractory metals (GPa). VRHZ stands for the Voigt-Reus-Hill averages and Zener anisotropy, Eqs.~(\ref{eq:bulk}-\ref{eq:aniso}). Stability violations are in bold red.}
\begin{tabular}{l|rrrr|rrrr|rrrr}
\hline
moduli &$C_{11}$&$C_{12}$&$C_{44}$&$\mu$ &$C_{11}$&$C_{12}$&$C_{44}$&$\mu$ &$C_{11}$&$C_{12}$&$C_{44}$&$\mu$ \\
VRHZ&$K_H$&$G_H$&$\sigma$&$A_Z$             &$K$&$G$&$\sigma$&$A_Z$             &$K$&$G$&$\sigma$&$A_Z$ \\
\hline
\hline
element &\multicolumn{4}{c|}{Ti}&\multicolumn{4}{c|}{V}&\multicolumn{4}{c}{Cr} \\
\hline
moduli  & 95&115&   41&{\color{red}\bf -10}   &278&143&  24&68        &444& 62&  99&191 \\
VRHZ   &108&{\color{red}\bf-10}&{\color{red}\bf 0.55}&{\color{red}\bf -3.97}  &188& 37&0.41&0.36      &188&129&0.22&0.52 \\
\hline
\hline
element &\multicolumn{4}{c|}{Zr}&\multicolumn{4}{c|}{Nb}&\multicolumn{4}{c}{Mo} \\
\hline
moduli  & 87& 93&   34&{\color{red}\bf -3}     &247&137&  17&55       &469&159& 101&155 \\
VRHZ   & 91&  5&0.47&{\color{red}\bf -10.4}    &173& 28&0.42&0.31      &263&120&0.30&0.65\\
\hline
\hline
element &\multicolumn{4}{c|}{Hf}&\multicolumn{4}{c|}{Ta}&\multicolumn{4}{c}{W} \\
\hline
moduli  & 73&116&   53&{\color{red}\bf -22}    &268&161&  79&54       &519&199&141&160 \\
VRHZ   &102&{\color{red}\bf -54}& {\color{red}\bf 0.83}&{\color{red}\bf -2.50}   &197& 67&0.35&1.45      &306&149&0.29&0.88 \\
\hline
\end{tabular}
\end{table}

Table~\ref{tab:elements} presents our results for pure elements of the transition metal columns 2-4 (starting with Ti, V and Cr).  Notice the stability violations for BCC structures of the Ti column, verifying their shear instabilities ($\mu < 0$) at low temperature.  Moduli and elastic stabilities increase, and Poisson ratios decrease, from left to right in the periodic table as nuclear charges and chemical bonding strength increase, and atomic volumes drop.

Moduli of our four BCC alloys are given in Table~\ref{tab:alloys}.  Notice that pair correlation function peak widths (see Fig.~\ref{fig:pdfs}) vary inversely with elastic moduli reported in Table~\ref{tab:alloys}; HCP/BCC alloys have the the broadest peaks and the lowest moduli.  These were obtained by applying a complete set of cell distortions while relaxing the atomic coordinates within the distorted cells, then taking two point central differences.  VASP was run on a GPU to accelerate the calculations~\cite{Hutchinson12,Hacene12}.  We employed $3\times 3\times 3$ $k$-point grids and an energy cutoff of 400 eV.  Three independent 128-atom configurations were utilized for each value, resulting in 9 independent measures of $C_{11}$ and $C_{44}$ and 18 measures of $C_{12}$ ({\em i.e.} taking $C_{22}$ and $C_{33}$ as independent values of $C_{11}$, etc.).  Standard deviations of the moduli were 2 GPa or less in every case.

\begin{table}
\caption{\label{tab:alloys} Elastic moduli (GPa) of BCC refractory HEAs.}
\begin{tabular}{l|rrrr|rrrr}
\hline
moduli &$C_{11}$&$C_{12}$&$C_{44}$&$\mu$
       & $C_{11}$&$C_{12}$&$C_{44}$&$\mu$ \\
VRHZ  &$K_H$&$G_H$&$\sigma$&$A_Z$             
       &$K_H$&$G_H$&$\sigma$&$A_Z$ \\
\hline
\hline
alloy  &\multicolumn{4}{c|}{NbTiVZr}&\multicolumn{4}{c}{CrMoNbV} \\
\hline
moduli  &161&103&   29&  29         &354&143&  51&105             \\
VRHZ    &122& 29& 0.39& 0.99        &213& 69&0.35&0.48           \\
\hline\hline
alloy   &\multicolumn{4}{c|}{HfNbTaZr}&\multicolumn{4}{c}{MoNbTaW}\\
\hline
moduli  &159&108&   41&  25         &371&160& 69& 106             \\
VRHZ    &125& 34& 0.38& 1.63        &230& 82&0.34&0.65 \\
\end{tabular}

\caption{\label{tab:averages} Averaged moduli of BCC refractory elements as defined in the text (Eqs.~(\ref{eq:Cbar}) and~(\ref{eq:bulkbar}).}
\begin{tabular}{l|rrrr|rrrr}
\hline
moduli &$\bar{C}_{11}$&$\bar{C}_{12}$&$\bar{C}_{44}$&$\bar{\mu}$
       &$\bar{C}_{11}$&$\bar{C}_{12}$&$\bar{C}_{44}$&$\bar{\mu}$ \\
VRHZ   &$\bar{K}_H$   &$\bar{G}_H$   &$\bar{\sigma}$&$\bar{A}_Z$     
       &$\bar{K}_H$   &$\bar{G}_H$   &$\bar{\sigma}$&$\bar{A}_Z$   \\
\hline
\hline
alloy  &\multicolumn{4}{c|}{NbTiVZr}&\multicolumn{4}{c}{CrMoNbV}  \\
\hline
moduli  &177&122&   29&  27         &360&125&  60&117             \\
VRHZ    &134& 19& 0.43&1.05         &201& 65&0.35&0.52            \\
\hline\hline
alloy   &\multicolumn{4}{c|}{HfNbTaZr}&\multicolumn{4}{c}{MoNbTaW}\\
\hline
moduli  &169&127&   45& 21          &376&164&  84&106             \\
VRHZ    &133& 14& 0.45&2.13         &229& 76&0.35&0.80  \\
\end{tabular}
\end{table}

To compare the moduli of HEAs with the individual elements, we present composition-weighted average elastic constants,
\begin{equation}
\label{eq:Cbar}
\bar{C}_{ij}=\sum_\alpha x_\alpha C_{ij}^{\alpha},
\end{equation}
in Table~\ref{tab:averages} ($x_\alpha$ is the mole fraction of chemical species $\alpha$).  In almost all cases, moduli of the HEA fall up to 20\% below the averaged elemental moduli, which we can understand as a reflection of the intrinsic disorder of the HEA.  The sole exception is the value of $C_{12}$ for CrMoNbV. Generalizing the Voigt, Reuss and Hill approach, we apply the rule of mixtures to define the isotropic moduli
\begin{equation}
\label{eq:bulkbar}
\bar{K}_V=\sum_\alpha x_\alpha K_H^\alpha,~~
1/\bar{K}_R=\sum_\alpha x_\alpha/K_H^\alpha,~~
\bar{K}_H=(\bar{K}_V+\bar{K}_R)/2
\end{equation}
and similarly for $\bar{G}$.  Notice that we are averaging elemental values of Hill moduli, $K_H$ and $G_H$, because the individual elemental crystalline grains are presumed randomly oriented.  We employ $\bar{K}_H$ and $\bar{G}_H$ in Eqs.~(\ref{eq:Poisson}) for the Poisson ratio $\bar{\sigma}$, however the anisotropy is taken as $\bar{A}_Z=2\bar{C}_{44}/(\bar{C}_{11}-\bar{C}_{12}$).  

\section{Atomic displacements}

Atoms displace from their ideal lattice sites due to thermal vibrations and random placement of differing sizes.  Sometimes referred to as ``lattice distortion''~\cite{Rosenhain23,Burgers46,Yeh06}, this effect reflects the interplay of thermal and interatomic forces creating displacements with elastic properties that resist them.  It is important because it can provide a mechanism to strengthen the alloy~\cite{Varvenne16,Oh16}.

We choose to define the lattice distortion, $\Lambda$, as the root-mean-square atomic displacement from the ideal lattice sites.  This choice is advantageous because provides the isotropic atomic displacement parameter~\cite{Trueblood96}, $U_{eq}$ via
\begin{equation}
\label{eq:LD}
\Lambda\equiv\sqrt{\avg{|\bu|^2}},~~~U_{eq}=\Lambda^2/3,
\end{equation} 
where the average is taken over atoms and time.  Thermally excited atomic vibrations generate the dynamical Debye Waller factor that scales diffraction peak intensities as $I(\bG)=I_0(\bG) e^{-2W}$.  In the case of isotropic elasticity, $2W=U_{eq}|\bG|^2$~\cite{Kittel}.

\begin{figure}
\includegraphics[width=4in, angle=-90]{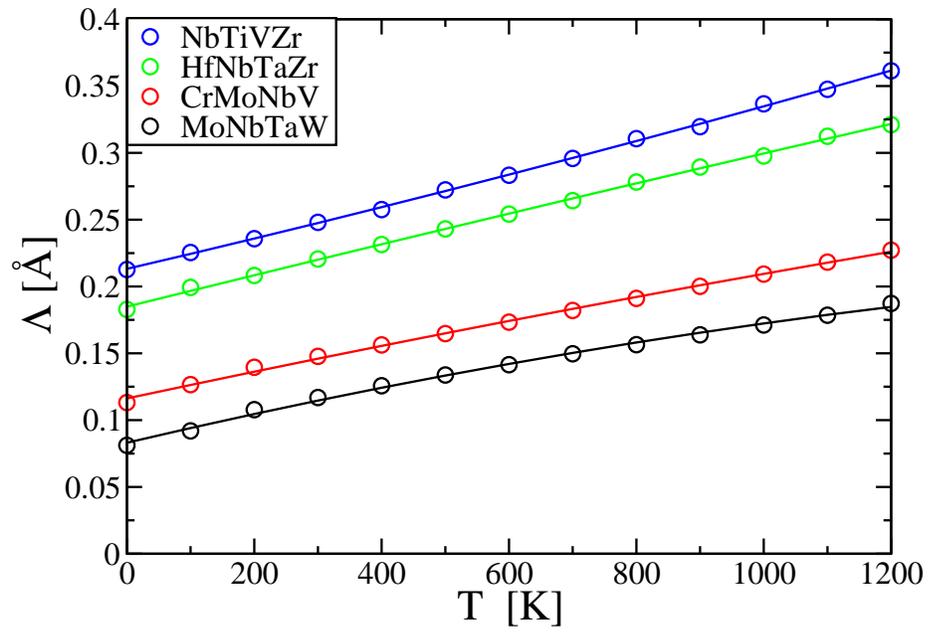}
\caption{\label{fig:Lambda} Lattice distortion $\Lambda(T)$ including size effect and thermal displacements (defined in Eq.~(\ref{eq:LD}).  Averages are taken over atoms and time, and over multiple independent quenches from MCMD configurations at $T=1200$K.  Calculations utilized a single $k$-point and default energy cutoffs.  Solid lines are quadratic fits.}
\end{figure}

Random interatomic interactions in our high entropy alloy create an additional static contribution to the lattice distortion that remains even as $T\rightarrow 0$ (see Fig.~\ref{fig:Lambda}).  This is known as the size effect~\cite{Krivoglaz,Schweika98,Welberry04} and its contribution to the Debye-Waller factor and to diffuse scattering were analyzed by Huang~\cite{Huang1947} in the limit of dilute solutions. According to this theory, the $T=0$K lattice distortion $\Lambda_0$ should grow proportionally to the variance of atomic size $\sigma_d=\sqrt{\avg{d^2}-\avg{d}^2}$.  Here we define the effective atomic size $d_\alpha$ of species $\alpha$ in the HEA as the peak position of the near-neighbor pair correlation function $g_{\alpha\alpha}(r)$.  Lattice distortion and size variance are compared in Table~\ref{tab:DW}.  Although $\Lambda$ varies monotonically with $\sigma_d$, they are not directly proportional.  This may reflect the complexity of concentrated solid solutions, where correlations in chemical occupation can influence the effective size of individual atoms.  For example, an excess of small atoms surrounding a large one may partially cancel the lattice dilation due to the large atom, but an excess of large neighbors could increase it.

Notice in Fig.~\ref{fig:pdfs} that the HEAs containing only normal BCC elements exhibit a preference for unlike (LR) neighbors.  This means that small atoms preferentially cluster around large, and vice-versa.  As a result, the static contributions to the $T=0$ lattice distortions are partially canceled by correlations, and the atoms remain close to their ideal lattice sites.  In contrast, for HEAs containing HCP/BCC elements, there is no such cancellation.  Indeed, there is a tendency for Zr atoms to segregate~\cite{Maiti16}.  As a result the static contributions may even be {\em enhanced} by correlations, and the atoms displace far from their ideal sites.

\begin{table}
\caption{\label{tab:DW} Comparison of $T=0$K lattice distortion ($\Lambda_0$, units~\AA), atomic size variance($\sigma$, \AA), slope of thermal displacements (${\rm d}U_{eq}/{\rm d}T$, \AA$^2$/K$\times 10^{-7}$) and predicted slope ($k_B/Ca$, units \AA$^2$/K$\times 10^{-9}$).}
\begin{tabular}{l|rrrr|rrrr}
\hline
        & $\Lambda_0$  & $\sigma_d$   & ${\rm d}U_{eq}/{\rm d}T$ & $k_B/Ca$
        & $\Lambda_0$  & $\sigma_d$   & ${\rm d}U_{eq}/{\rm d}T$ & $k_B/Ca$ \\
\hline
\hline
alloy   &\multicolumn{4}{c|}{NbTiVZr} &\multicolumn{4}{c}{CrMoNbV} \\
\hline
        &0.213 &0.341  &  151&  658   &0.113&0.239&  81&307          \\
\hline\hline
alloy   &\multicolumn{4}{c|}{HfNbTaZr}&\multicolumn{4}{c}{MoNbTaW}\\
\hline
        &0.183 &0.314  &  147&  556   &0.081&0.173&  70&243          \\
\end{tabular}
\end{table}

Thermal contributions to the atomic displacement parameter can be evaluated from the phonon dispersion relation $\omega_s(\bk)$, where $\omega$ is the frequency of a mode of polarization $s$ ({\em e.g.} longitudinal or transverse) and $\bk$ is the phonon wavevector.  In the continuum limit at long wavelengths, $\omega_s(\bk)=v_s K^2$ where the sound speed $v_s=\sqrt{C_s/\rho}$ with $C_s$ a suitable combination of elastic moduli, and $\rho$ the mass density. For example: $C_s=C_{11}$ or $C_{44}$, respectively, for longitudinal or transverse sound propagating in the cubic $[100]$ direction; $C_s=(C_{11}+2C_{12}+4C_{44})/3$ or $(C_{11}-C_{12}+C_{44})/3$ in the $[111]$ direction; $C_s=(C_{11}+C_{12}+2C_{44})/2$, $(C_{11}-C_{12})/2$, or $C_{44}$ in the $[110]$ direction~\cite{Kittel}.

Applying the law of equipartition for classical harmonic vibrations, we have
\begin{align}
\label{eq:ADP}
U_{eq} &= \frac{N}{3}\sum_s \int\frac{{\rm d}\bk}{V_{BZ}} \frac{k_BT}{\rho V \omega_s^2(\bk)} \\
\label{eq:ADPapprox}
       &\sim \frac{k_B T}{C a},
\end{align}
where $N$ is the total number of atoms in volume $V$.  We evaluated the approximation in Eq.~(\ref{eq:ADPapprox}) by assuming a Debye density of states with isotropic sound speeds, relating the Brillouin zone volume $V_{BZ}$ to the BCC lattice constant $a$, and dropping factors of order 1.  The elastic constant $C$ is defined as an average over orientations and polarizations {\em via}
\begin{equation}
\label{eq:C}
\frac{3}{C} = \sum_s \int\frac{{\rm d}\bkh}{4\pi} \frac{1}{C_s(\bkh)}.
\end{equation}
In practice we shall evaluate $C$ through a multiplicity-weighted sum over 2-fold, 3-fold and 4-fold symmetry directions.  $U_{eq}$ grows linearly with respect to temperature $T$, provided we remain in the harmonic approximation assumed in Eq.~(\ref{eq:ADP}).  Table~\ref{tab:DW} tests our prediction (Eq.~(\ref{eq:ADPapprox})) and shows that the thermal displacements vary monotonically with respect to $k_B/Ca$.  The lack of exact proportionality is due, at least in part, to our application of the isotropic Debye model in our derivation.

\section{Conclusions}

In conclusion, we demonstrate that inclusion of BCC/HCP elements in refractory HEAs reduces the shear modulus $\mu=(C_{11}-C_{12})/2$.  Indeed, a tetragonal distortion has been reported in the Zr-rich regions of phase segregated BCC/HCP containing HEAs~\cite{Maiti16}.  This effect may be related to the elastic instability of the BCC/HCP elements arising from the Fermi level lying in a peak of the electronic density of states.  Although inclusion of regular BCC elements stabilizes the BCC HEA, it should be possible to drive the system close to a shear instability by increasing the BCC/HCP content, and thereby achieving a high Poisson ratio and potentially enhancing ductility. Examples of compositions predicted to be on the threshold of low temperature instability based on the averaged elemental modulus $\bar{\mu}$ are Nb$_{0.06}$Ti$_{0.63}$V$_{0.06}$Zr$_{0.25}$ and Hf$_{0.53}$Ta$_{0.11}$Nb$_{0.11}$Zr$_{0.25}$.

We quantified the $T=0$K lattice distortion $\Lambda_0$ and demonstrated growth of $\Lambda(T)$ at $T>0$ is governed by the inverse elastic moduli.  At $T=0$ we find that the atomic size effect contributes to lattice distortion as predicted by Huang~\cite{Huang1947}, but we found evidence for other effects associated with interatomic correlations that may either diminish or enhance distortion depending on whether large and small atoms attract or repel.  Engineering alloys to alter interatomic interactions or to reduce shear moduli can thus enhance lattice distortion and thereby potentially increase hardness relative to undistorted structures with similar elastic moduli.

\section*{Acknowledgement}
This work was supported by the Department of Energy under grant DE-SC0014506 and by the Pittsburgh Supercomputer Center under XSEDE grant DMR160149. We acknowledge useful discussions with Michael Gao and Takeshi Egami.
\section*{References}

\bibliography{hea}

\end{document}